# The Quantum Master and its Classical Emissary

Ruth E. Kastner
"Metaphysics and the Matter with Things"
March, 2024

ABSTRACT. Ian McGilchrist's works present the thesis that the two hemispheres of the brain have radically different modes of interacting with the world, and that their respective perceptions and functions must be properly integrated for a viable way forward. This proper integration requires restoring the right-brain to its proper place as "Master." I discuss a parallel to this insight in the dichotomous "worlds" of quantum and classical physics. In addition, I discuss the relevance of Whitehead's process philosophy, as well as the Taoist concepts of Yin and Yang, with particular attention to the importance and primacy of Yin underlying the quantum level as "Master."

1. Introduction

Ian McGilchrist's works (McGilchrist 2019, 2024) lay out and elaborate the thesis that the two hemispheres of the brain have radically different views of the world and ways of interacting with the world, and that their respective perceptions and functions must be properly integrated for a viable way forward. This proper integration requires restoring the right-brain to its proper place as "Master." We can find a significant parallel to this insight in the apparently dichotomous "worlds" of quantum and classical physics. Roughly speaking, quantum theory describes the sub-microscopic realm of tiny entities remote in empirical availability, such as atoms and subatomic particles, while classical physics describes the macroscopic, empirical realm of everday experience.

Classical physics is characterized primarily by Newton's laws, which describe localized objects and deterministic processes. While Newton's laws have been surpassed in the form of Einstein's relativity theory, the description provided under classical physics is still that of what has come to be called "local realism." This term denotes apparently localized and separable objects, each with its own independent individuality, or "being thus," as Einstein termed it. The philosophical view closely aligned with this metaphysical picture is *nominalism,* which sees the universe as comprised solely of individualized entities. However, quantum theory presents us with features that were unexpected from the standpoint of nominalistic classical physics, starting with the theory's incompatibility with local realism. Quantum states, which describe the microscopic systems subject to quantum theory, "live" in Hilbert space, a complex multidimensional mathematical space that is decidedly not the 3+1 real-valued parameters of space and time. This feature of the quantum formalism leads to the various peculiarities for which quantum theory has become notorious.

Among these peculiarities are the Uncertainty Principle and quantum entanglement. The Uncertainty Principle, first elucidated by Heisenberg, prevents an object from having the

"complete set" of properties habitually presupposed by Western science. For example, it precludes an object's possession of both a definite position and momentum at any given time. Quantum entanglement leads to violation of the locality of objects and information transfer that has long been thought to be a requirement for a sensible physical account of reality. For example, two entangled electrons comprise a whole that is in a precise mathematical sense "greater than the sum of its parts," such that measurements on each of the electrons must be instantaneously coordinated in a way that appears to violate locality, or at least to deny the classical idea that each electron has its own, independent 'being thus.'

A third peculiarity is the so-called "Measurement Problem," which is arguably a deficiency of the theory in its conventional form (cf. Kastner 2023). The measurement problem consists in the inability of conventional quantum theory to say what sort of interaction counts as a "measurement" leading to an outcome that can be described by the corresponding quantum state. The measurement problem is famously illustrated by the Schrodinger's Cat Paradox, in which a quantum superposition of an unstable atom, whose time of decay is uncertain, seems (according to the conventional theory) to propagate up to the macroscopic level, leaving a cat in a superposition of "alive and dead". In reality, we never see macroscopic objects like cats in superpositions, since one or the other outcome has already occurred at a level below the macroscopic.[1] Thus, while the conventional theory provides well-corroborated probabilies for the outcomes of measurements, it fails to say how or why we actually *get outcomes*. This issue, while not our main focus, will be briefly revisited later in Section 3 when we consider the instructive parallel between the Yin/Yang duality of Taoism and the Right/Left duality of the brain hemispheres. In that section we will note that the conventional theory is missing the dynamical "Yin" component, and that remedying this lacuna offers a solution to the measurement problem.

The overall thesis of the present work is that the quantum level is rightfully regarded as the fundamental "Master" of physical reality, while the classical level is a secondary "Emissary" that arises under specific circumstances. In effect, the classical level functions as a mere "user interface" between an observer and the totality of physical reality. In beginning to explore this notion, let us recall the different modes of knowing and operation that McGilchrist has laid out, as symobolized in Figure 1:

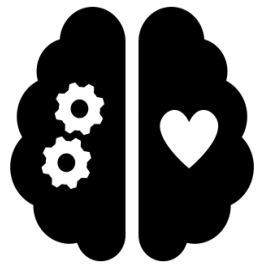

Figure 1. The left brain operates in an analytic, mechanistic mode, while the right brain operates in a holistic, intuitive mode.

---

[1] While it is often claimed that 'decoherence' solves this problem, it does not actually do the job. This issue is discussed in Kastner (2014) and Kastner (2020).

As McGilchrist has noted, the left hemisphere of the brain employs analytic modes of thought in which sensory impressions are categorized into apparently separate objects, each pursuing its own local and deterministic trajectory. Thus, the left-brain works with a *mechanical* picture in which a given phenomenon is localized in space and projected into the future based on its remembered past behavior. This mode also abstracts and generalizes to impose mechanistic rules on its perceived reality. In contrast, the right hemisphere does not attempt to localize phenomena nor map them "into" space and time this way, but is attuned to the present moment, and has a global awareness of all that is present in that moment. This awareness has an intuitive and synthetic aspect that transcends the mechanical analytic left-brain mode and is not concerned with rules or constraints. That is why it is often stated that the right-brain is the seat of creativity.

Returning now to the physical theories mentioned above, recall that the classical level is empirically available and appears as concrete, localized objects undergoing separable processes, while the quantum level is pre-empirical in that it is not directly available to the external five senses. The quantum level also has strange, nonlocal features that seem to contradict what we are used to in our observed phenomena. We can visualize this distinction in terms of an "iceberg," where the tip is described by classical physics and the submerged portion is described by quantum physics:

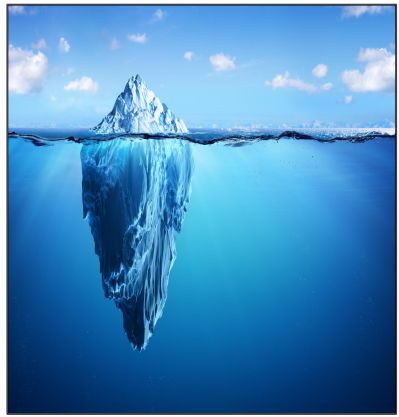

Figure 2. Classical physics describes the 'tip of the iceberg', while quantum physics describes the submerged, unseen portion.

Classical physics thus corresponds to a left-brain type of understanding, while the quantum level, with its holism, indeterminacy, and nonlocality, corresponds to a right-brain type of understanding. Thus, the right brain can be seen as 'tuned into' the quantum level, while the left brain is only able to deal with the classical level.

We now observe that much of modern thinking is dominated by left-brain modes of thought, such as either/or categories ("law of excluded middle") and mechanistic conceptualization. Thus, the left-brain 'emissary' is inappropriately in charge. This leads to the current situation in which the classical level--with its fragmented, mechanistic quality--is generally mistaken for the "true" reality, and prevailing attempts to come to grips with the implications of quantum physics for our reality are hampered by left-brain constraints. Thus, modes of understanding that are more aligned with right-brain processes are needed in order to recognize the quantum level as the fundamental reality, and to successfully account for its manifestation at the classical, empirical level.

2. Manifest vs. Unmanifest Reality

Let us now recall a famous quote from psychologist Carl G. Jung :

*"The multiplicity of the empirical world rests on an underlying unity…everything divided and different belongs to one and the same world, which is not the world of sense." –Carl Jung, Mysterium Coniunctioni (1977).*

This remark seems to coincide quite strikingly with the identification we have made above, namely: the empirical world is the realm of classical physics that describes everyday phenomena, which apparently reflect distinct and separated objects pursing deterministic careers. On the other hand, the quantum level is not directly available to the external senses and and has an undivided, nonlocal, holistic character. D.V. Ponte and Lothar Shäfer are among those who have remarked on this idea of the quantum level as a hidden domain underlying our everyday world of appearance. They have this to say about Jung's body of work:

"If we want to characterize Carl Jung's psychology in one sentence, we can say [that it] leads us to the view that there is a part of the world that we can't see, a realm of reality that doesn't consist of material things but of non-material forms."   -- Ponté and Shäfer (2013)

While the characterization of "non-material" may not be quite apt (or at least, is a matter for further discussion), Ponte and Shäfer point to the idea that the quantum level is more abstract, more mind-like than the phenomenal level shown to us by our external senses. This observation can be framed in terms of the distinction between (1) the manifest level of actuality and (2) the unmanifest level of potentiality (cf. Kastner, Kauffman and Epperson 2018). These categories can be summed up as follows:

| **Unmanifest** (submerged portion of iceberg) | **Manifest** (tip of the iceberg) |
|---|---|
| "Quantum" | "Classical" |
| unactualized; potential | actualized |
| wavelike/organic | particle-like/mechanical |
| nonlocal correlations | local behavior |
| holistic; a unity | apparently fragmented |
| "becoming" framework | "Factual" framework |
| pre-spatiotemporal | spacetime level |

In order to realize this as an integrated picture, it's helpful to recover the overlooked role of

"Yin" in Western thought. We turn to this issue in the next section.

   3. The Relevance of Taoist Concepts: Yin and Yang

   Let us first recall the Taoist concept of Yin and Yang, a complementary pair of interactive processes that are seen as comprising a unity that transcends their apparent duality. The Yin and Yang duality and its significance are expressed in the Taoist sage Lao-Tzu's timeless book of aphorisms, the *Tao Te Ching*. The concept is represented by a well-known symbol, the *Taijitu*:

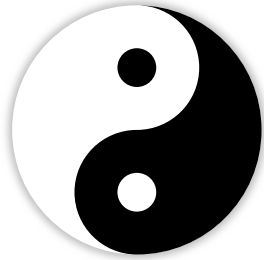

Figure 3: The Taijitu representing the toality of Yin and Yang

   The general characteristics of Yang and Yin are exemplified in the following table. There are many other examplars of each; we offer just a sample here.

| Yang | Yin |
| --- | --- |
| light | dark |
| initiating | responding |
| creating | destroying |
| offering | receiving |
| emitting | absorbing |

   We can see in the symbol of Figure 3 that each element contains the "seed" of its opposite. The Yin/Yang duality is a dynamic one of process, in which one or the other may appear to be dominant at any point around the circle, but the totality is a "dance" of which both elements are constitutive, and flow into one another to create a unified process.

   It is often mistakenly assumed that only Yang is active, in a false dichotomy of "action versus inaction" or "activity versus passivity," but this is a misleading caricature of the Yin/Yang duality. While the Yang-type processes are obviously active in nature, the Yin-type processes, while subtler and seeming secondary, are decidely neither passive or inactive. For example, someone who wants to sell their home (offer, Yang process) must have an active buyer (receiver, Yin process) in order to accomplish their goal. Another example is found in reproductive biology, which was traditionally viewed as "all Yang" in that the male sperm was considered the only active element. It wasn't until the later part of the 20th century that researchers seriously studied the process and discovered that the ovum is an active partner that sends out responding chemical signals to help sperm find their way, and even generates a "fertilization cone" that envelops the inital sperm and prevents others from gaining access. The ovum's Yin-role is not only active, but crucial to fertilization. Emily Martin (1991) famously pointed out this traditional neglect of the egg's active role in terms of gender bias in the sciences, but it also demonstrates the conventional exclusion of Yin-type processes, which have not been "seen" by Western science.

Nevertheless, there is an inherent asymmetry between Yin and Yang, and there are important reasons to see it the opposite of the prevailing attitude described above. Ironically (for the Western mindset), Yin may be properly regarded as the more powerful and fundamental. This has long been a feature of Eastern thought. It is expressed in Verse 28 of the *Tao te Ching,* which begins with the statement: "Know the Yang, but keep to the Yin." The fundamentality of Yin is elaborated as Verse 28 continues: "Know the bright, but keep to the dark...Thus become the abundant valley of the world." In fact, in a sense the Tao itself (a totality of existence, roughly translated as "the Way") is seen as far more Yin-like than Yang-like, as expressed in Verse 4: "The Tao is like a well; used but never used up. It is like the eternal void: filled with infinite possibilities. It is hidden but always present." Verse 11 continues with the practical utility of Yin: "We shape clay into a pot, but it is the emptiness inside that holds whatever we want." This theme---that the background and receiving aspects of reality are primary---is stressed throughout the *Tao Te Ching*. Quantum theory provides an additional reason to view Yin as more fundamental when the theory is understood as pointing us to a hidden, far larger, submerged portion of the metaphorical iceberg--an enormous domain of unmanifest possibilities underlying the actualized "tip of the iceberg" phenomenal reality we directly perceive. In a real sense that goes beyond metaphor, the quantum realm of possibility is a kind of 'valley,' or darkness from which light arises.[2]

Recall that McGilchrist's thesis is that the right-brain should be regarded as Master, having a function and role of oversight for the left-brain who is the Emissary. Identifying Yin with the right-brain reflects the above-discussed intrinsic asymmetry between Yin and Yang that places Yin in the "Master" role. We can thus augment the table, indicating that Yin is to be associated with the Unmanifest and also with the right-brain modes of thought:

| **Unmanifest** (submerged portion of iceberg) | **Manifest** (tip of the iceberg) |
| --- | --- |
| "Quantum" | "Classical" |
| unactualized; potential | actualized |
| wavelike/organic | particle-like/mechanical |
| nonlocal correlations | local behavior |
| holistic; a unity | apparently fragmented |
| "becoming" framework | "Factual" framework |
| pre-spatiotemporal | spacetime level |
| domain of Yin | domain of Yang |
| right brain | left brain |

---

[2] This is quite literally the case in the transactional formulation, in which photons are generated from the hidden quantum realm and serve to create the structured set of events that comprise what we call "spacetime." Technical details are found in Kastner (2022), Chapter 8 and Schlatter and Kastner (2023). It may be noted that a "black hole" can be understood as a "window" into the unmanifest quantum level.

4. Western thought: All Yang, No Yin

As alluded to above, the prevailing Western left-brain framework puts us in a constraining metaphysical box characterized by an overdependence on Yang-like characterizations and neglect of the crucial Yin processes. In effect, the classically-restricted Emissary is in charge instead of the quantum-aware Master. This leads to the key aspects of the Western metaphysical paradigm:

1. Cartesian dualism: *res extensa* vs *res cogitans*, with no interaction between them ('mind-body problem'. Or its materialistic variant: only *res extensa* exists; mind is viewed as an epiphenomenon of *res extensa*
2. actualism
3. separate objects are fundamental (nominalism)
4. causal locality (field propagation is modeled as 'bucket brigade')
5. All Yang and no Yin (physical processes are assumed to be unilateral)

Features 1-3 are fairly straightforward given the preceding, but for clarity, let us elaborate 4 and 5, including their terminology. In a general sense, a *field* is an area of influence generated by sources (such as a charge generating an electric field). In conventional quantum field theory, a field is viewed as a "system of oscillators associated with every spacetime point" -- the latter being a typical textbook characterization. In other words, the field is modeled as an independently existing mechanical system that "carries" energy from one point to another in a local manner, in what is viewed as a spacetime "container" or background. The latter is presumed to delimit everything physically real.[3]

The conventional theory models energy propagation as involving an excitation that is launched by an emitting source and transmitted "through space and time" from one system to another by the postulated field oscillators, in a local manner. The receiving system is viewed as passive; i.e., it is assumed to contribute nothing to the creation of the transferred excitation. Besides its locality, this picture assumes that field propagation is directed unilaterally from one system to another, just as in the kind of bucket brigade of old-fashioned firefighting (depicted in Figure 4). The firefighters are analogous to the "field operators" corresponding to the hypothesized oscillators, while the buckets of water are the field excitations--energy quanta--generated by the emitter. An example of such an excitation is a photon, the "quantum of light." Thus, we can call this the "bucket brigade" picture of field propagation. Its key features are locality and uni-directionality.

---

[3] If spacetime is viewed as a realm of actuals, which is the conventional view under which potentiality is not considered physically real, then this picture implies #2, actualism.

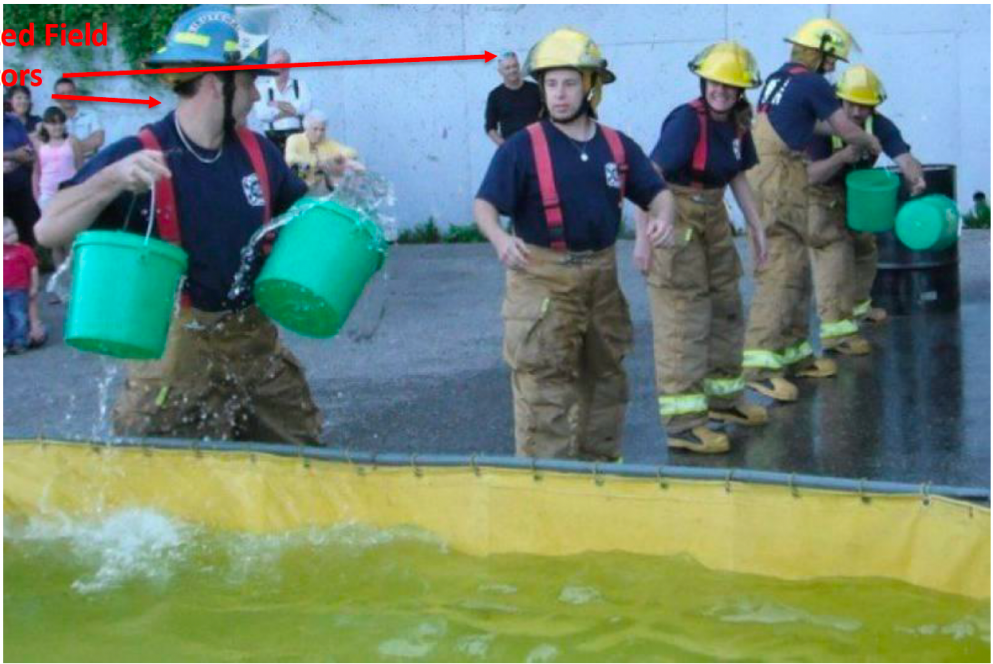

Figure 4. The "bucket brigade" picture of field propagation.

This concept of field propagation exemplifies a purely Yang-type process. It is viewed as an individually creative act initiated unilaterally--in this case, by the "emitter" using the tool at hand--the postulated mechanical oscillators--by exciting them and thereby causing them to deliver its desired message, which is dumped into the final sink (water tub). The latter is seen as passive and contributing nothing to the process. This all-Yang notion of field propagation mimics the kinds of macroscopic phenomena we see at the usual empirical level, such as tennis serve (Figure 5).

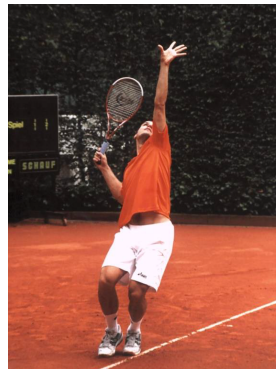

Figure 5. A tennis serve as a Yang-only process

So, let us sum up the key features of the "Yang-only" metaphysics underlying conventional Western science (and in particular, physics), with the help of the "tennis serve" analogy:

- Unilateral (All Yang: the 'tennis serve' picture)
- Nothing happens unless a single player hits the ball
- Then everything happens (i.e., nobody needs to hit or catch the ball for it to pursue a trajectory)
- The ball pursues an autonomous trajectory as an independent object
- The naive classical view of localized energy propagation---a 'bucket brigade'

Again, these features may seem obvious and even necessary based on our ordinary macroscopic experience. However, they are arguably constrained by left-brain perceptual and conceptual limitations, and their status as unexamined background assumptions, uncritically imported into theoretical descriptions of the microscopic level, leads to various problems in physics. Chief among these is the Measurement Problem of quantum theory, which consists in the inability of the conventional theory to specify what sort of interaction counts as a 'measurement' yielding an outcome. In what follows, we consider how proper integration of the Yin element in quantum theory can resolve this issue and shed light on the various perplexities of the theory.

5. Quantum perplexities and their resolution through Yin and right-brain modalities

As mentioned above, quantum theory presents us with various perplexing conceptual challenges. We recall the key issues here:

1. The Measurement problem ("Schrodinger's Cat" Paradox)
2. Uncertainty (Heisenberg Uncertainty Principle; "contextuality" of properties)
3. Entanglement & Nonlocality

Left-brain modes of thought lead to the Yang-dominant theoretical limitations resulting in (1), a defect of conventional quantum theory. They also lead to aversion to (2) and (3), which are natural features of the Yin-like quantum realm. The remedy for this situation is to relinquish the Yang-only metaphysics afflicting classical habits of thought. The key step in doing so is to incorporate a long-available but historically neglected formulation of field behavior, namely the so-called "absorber theory of radiation," also known as the "direct-action theory of fields" (DAT). This theory is the basis of the Transactional Interpretation of quantum theory (TI). While the details of this formulation are beyond the scope of this work, we can summarize its key features as follows:[4]

1. The Transactional Interpretation: includes Yin (responses of field sources)

---

[4] The interested reader can find introductory accounts of TI in Kastner 2015, 2016, 2022.

2. Really a different formulation of quantum theory: based on the 'absorber' theory of radiation (also known as the 'Direct-Action Theory' or DAT).
3. TI rejects these classical Western Yang-only assumptions:
   unilateral propagation ("tennis serve")
   locality (information/influences always propagate at or below the speed of light)
   influences always localized at some spacetime point

As noted above, TI uses the direct-action theory of fields, in which field sources (e.g., charged particles such as electrons) generate time-symmetric fields of influence that are intrinsically nonlocal and non-unilateral. These interactions are irreducibly *mutual and relational* in that they always join a set of field soures such that there is no matter of fact about which sources 'generate' the field and which 'receive' the field (Figure 6).[5] These field connections are called "virtual photons." They mediate electromagnetic force, but do not rise to the level of radiated energy.

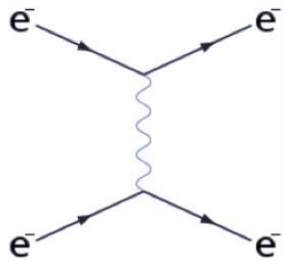

Figure 6. the basic time-symmetric field of the direct-action theory, a virtual photon.

However, under appropriate conditions (e.g., a field source in an excited state and satisfaction of the conservation laws for quantities of energy, momentum, etc. transferring from an emitter to an absorber), this mutual relationship changes in character to an offer/response situation. Specifically: *both* (or all) participating systems generate their own time-symmetric field such that the excited source becomes the offering emitter and others become responding absorbers (see Figure 7).[6] The emitter's field is called an "offer" and the absorbers' responding fields are called "confirmations." The combined fields create a quantum of electromagnetic energy---a real photon--that is transferred from the emitter to one of the responding absorbers. This process is called a "transaction":

---

[5] Interestingly, even in conventional quantum electrodynamics (QED), a "virtual photon" is explicitly represented as a connection in which there is no fact of the matter about which field source emits and which receives (this is the Feynman propagator). However, the conventional theory cannot physically distinguish a virtual photon from a real photon: real photons are just modeled as "external lines" rather than "internal lines" and it is simply stipated that they are emitted in one location and absorbed in another, based on the observed empirical detections. In contrast, the direct action theory disambiguates the virtual from the real photon; they are crucially distinct both in physical nature and function. See Kastner (2022), Chapter 5, for further details.
[6] This process was called "absorber response" in the original version of TI (Cramer, 1986), but it is really a mutual process involving both the emitter and the absorbers involved in the interaction; see Kastner (2022), Chapter 5 for details.

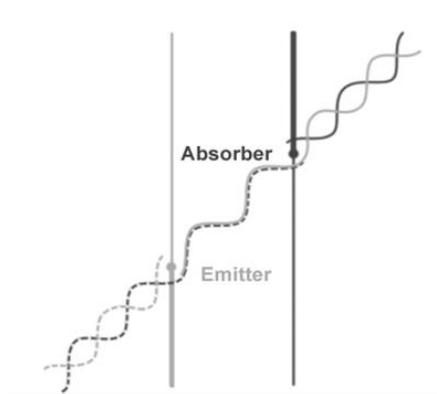

Figure 7. A transaction, in which a photon is radiated from an emitter to an absorber.

    The transactional process involves both spatial and temporal symmetry breaking, such that the photon is delivered *from* the source *to* one of the reponding absorbers in what is called a "radiative" process. This involves "quantum collapse," since in general more than one absorber responds, yet there is only one photon (or some limited number) available for transfer. Many absorbers play a part in the interaction, yet only one of them can actually receive a single photon.[7] The delivery of the photon to one of the participating absorbers constitutes an *actualized transaction*. The latter constitutes a newly emergent element of the spacetime manifold or 'tip of the iceberg.' The photon can be understood as the structural link between a newly established *emission event* and *absorption event*, which constitute an indivisible pair--one never occurs without the other. This linked pair can be seen as a basic element of *res extensa*, which is emergent rather than primary in this picture.[8] Specifically, it emerges from a quantum substratum or domain of *res potentia* (Kastner, Kauffman, Epperson 2018).

    We can see both Yang (offer) and Yin (response) components in the transactional process. And it turns out to be the occurrence of the Yin response that allows one to define the process of "measurement" that is missing the conventional Yang-only theory. Specifically, "measurement" occurs upon generation of responding fields by absorbers and the transfer of one or more real photons--a transaction, which involves *both* Yang- and Yin-type processes. In the conventional Yang-only theory, one cannot make this kind of quantitative distinction between virtual and real photons--all interactions are assumed to be of the same kind, of the "tennis serve" type, so there can be no transaction (and thus no measurement outcomes). The Yin process of response, crucial for radiation in the direct-action theory, is entirely missing in the conventional account. As a result, the conventional theory can provide no distinction between interactions that do not result in a measurement outcome and those that do. And that is the Measurement Problem, which is remedied in the transactional formulation that includes both Yin and Yang.

---

[7] This point also accounts nicely for the phenomenon of "null measurement," when it is known that a measurement occurred but a particular detector did not fire. That detector is involved in a null measurement: it responded to an emitter but did not "win" the photon.

[8] The interested reader is invited to consult Kastner (2022), Chapter 8 for details of this picture of spacetime emergence from the quantum level. Further, it leads to a quantum theory of gravity that is transactional in nature, as presented in Schlatter and Kastner (2023). This theory naturally yields the effects conventionally attributed to 'dark matter' and 'dark energy'.

This picture also involves relinquishing the classical idea that spacetime is a "container" or background for everything that exists, and that all processes must be defined relative to that background in a localized and separable manner. That is essentially local realism, which arises from left-brain modes of analysis. But the quantum level defies this expectation; quantum field sources (charges and collective states of charges, such as atoms) are, in general, non-separable (this is entanglement). They demand multidimensional and complex states, in keeping with their ontological status as elements of *res potentia* in the quantum substratum; this corresponds to property indefiniteness quantified by the uncertainty principle. And in the direct-action (transactional) picture, the field connections are intrinsically nonlocal. While the left-brain vehemently rejects this ontology, the right brain has no problem with it. Thus, we see that allowing the right-brain to be "in charge," and allowing Yin elements (both dynamical and ontological) their due role, the quantum peculiarities listed above are either resolved or recognized as mere aversions of the left-brain.

6. Whitehead, Prehension, and Yin

The above picture of field behavior has significant features in common with Alfred North Whitehead's *process philosophy*. While space limitations preclude a full examination of this issue in the present work, we can note the salient features. The key relevant aspect is Whitehead's notion of *prehension*, which is seen as a generative feature of the dynamical process of becoming that underlies our phenomenal universe. Prehension means "grasping," or "taking into account," and is decidely a Yin-type process. Of course, something needs to be "there" already in order to be grasped and taken into account, so that Yang is involved; yet the Yang 'offer' or created element still needs to recognized and actively responded to/accepted, grasped. In Whitehead's formulation, prehension is a fundamental aspect of the establishment of the *occasions of experience* that are the happenings, or elements of the process of becoming as he conceived it. And we saw in the previous section that it is an actualized transaction, which consists of both an offer and a confirmation---the latter being a form of prehension---that yields an actualized "link" or element of the emergent spacetime fabric. This element can be identified with Whitehead's occasions of experience, or *actual entities*. A Whiteheadian actual entity is just an actualized transaction, a primary element of spacetime. And prehension, in a real physical sense, plays a crucial role in this causal process of spacetime emergence.[9]

Significantly, prehension is a form of behavior that is not part of conventional theoretical modeling in physical science. Conventional physical science sees and accepts only conceptualizations in terms of generative, extending, and persisting concepts--Yang behaviors.

---

[9] This picture arguably sheds light on the elusive nature of causation. As Hume noted, we do not detect causation empirically as "sense-data"---what Whitehead would call *perception as presentational immediacy*. Hume chose to discount causation on that basis, while Whitehead retained it, allowing for a broader range of modes of perception, including *perception as causal efficacy*, and noting that "the ingression of objects into events includes the theory of causation" (Whitehead 1920, 146). In the transactional formulation, the relevant "objects" are the transacting fermionic quantum systems, and "events" are the emissions and absorptions in which they take part. We thus see that causation, involving "ingression of objects into events," is a physical process taking place at the quantum pre-spatiotemporal level, involving creation of the very photons needed to activate our sense organs. To detect a photon on the retina is to empirically experience only the final result or completion of that causal process.

Yet without Yin, nothing can actually happen, because nothing can be recognized and accepted--nothing can be prehended. (Can anyone sell their home without an active buyer who recognizes and chooses to accept it?) Thus, prehension, a Yin-process, remains crucial. Yet it is either not seen or actively eschewed in standard scientific and philosophical theorizing. This methodological antipathy to Yin may perhaps be due to its arguable inseparability from *perception*:[10] something must be perceived in order to be taken into account, and that seems to the materialistic mind as an illicit smuggling in of subjectivity or even animism.[11] That is, the conventional scientific mindset is based on the Cartesian notion of nonliving substance---pure *res extensa,* taken as primary. A nonliving substance certainly cannot "take into account" anything. The dead do not keep accounts.

Thus, Whiteheadian prehension would appear to clash directly with Cartesian metaphysics, which takes "dead matter" or pure *res extensa* as the primary constructive substance of the world. In contrast, the picture laid out in the previous section includes *res extensa* only as a final result that supervenes on a process of prehensive becoming. We perhaps see in this situation the deep reason for the persistent marginalization of Whitehead's philosophical approach: namely, its repudiation of the mechanistic, Cartesian metaphysical grounding of conventional physical science. Yet the price paid for the longstanding conventional adherence to Cartesian "dead matter" is a dead end in attempts to account for "measurement" in quantum theory, as well as ongoing confusion and discomfort with the Yin-like aspects of quantum theory that arise from its Hilbert space structure, such as nonlocality and "contextuality" or property indefiniteness.

7. Conclusion: Quantum Level as (Yin) Master, Classical Level as (Yang) Emissary

In the foregoing, we have provided an overview of the insight that recognition of Yin-type processes and their appropriate inclusion in the dynamics of field interactions can resolve the notorious Measurement Problem of conventional quantum theory, as well as shed new light on some of the perplexities of quantum theory. We have also noted that Yin can be seen as more fundamental, as it naturally accommodates the quantum level's background-like, unmanifest, "hidden" aspects. In contrast, the Yang-like processes can be associated with the phenomenal and emergent, classical, spacetime level of ordinary experience, which can be seen as a kind of "user interface" between a perceiving person and the unseen quantum level. We have noted that these observations align naturally with the right-brain vs. left-brain modes of perception and thought as emphasized by the work of Iain McGilchrist. Thus, the advent of quantum physics--however resisted by the conventional left-brain mindset--can be seen as reinforcing Iain McGilchrist's proposal that right-brain modes of perception are more fundamental than those of the left-brain and belong "in charge": the quantum level is the "master," while the classical level is the "emissary."

---

[10] For our purposes, perception in the sense of causal efficacy is the relevant meaning (see previous note). This is distinct from the conventional Western idea of perception as involving only presentational immediacy, as in "sense data." The latter corresponds most closely to spatiotemporal events that are the only the completion of the process of becoming. See Gomes (2015) for an insightful discussion of the distinction between these two modes of perception. Gomes recalls that Whitehead viewed the causal efficacy mode as more fundamental, which agrees with our account here.

[11] That prehension suggests (and even implies) some form of subjectivity is argued by L. B. McHenry (1995).

We have also noted the close correspondence of this formulation with the Whiteheadian picture of process and becoming, in particular the importance of the Yin-like concept of prehension. We have also suggested that conventional avoidance of Whitehead's approach can be seen as arising from unexamined adherence to Cartesian metaphysics.

Acknowledgments. I would like to express my appreciation to the Center for Process Studies and the California Institute of Integrative Studies for the opportunity to participate in this conference celebrating the work of Iain McGilchrist. I also am grateful for Ronny Desmet for valuable comments on an earlier draft.